\documentclass[aps,pre,groupedaddress,twocolumn]{revtex4}
\usepackage{hyperref}
\usepackage{subfigure}
\usepackage{color}
\usepackage{graphicx}

\begin{document}
\expandafter\ifx\csname urlprefix\endcsname\relax\def\urlprefix{URL }\fi

\title{\large
 From the Newton's laws to motions of the fluid and superfluid vacuum:\\
 vortex tubes, rings, and others }

\large
\author{Valeriy I. Sbitnev}
\email{valery.sbitnev@gmail.com}
%\affiliation
\address{B. P. Konstantinov St. Petersburg Nuclear Physics Institute, NRC Kurchatov Institute, Gatchina, Leningrad district, 188350, Russia;\\
 Department of Electrical Engineering and Computer Sciences, University of California, Berkeley, Berkeley, CA 94720, USA
}

% Comments: 11 pages,  8 figures
% Subjects: Quantum Physics (quant-ph, physics.gen-ph, physics.flu-dyn)

\date{\today}

\begin{abstract}
 Owing to three conditions (namely: (a) the velocity is represented by sum of
 irrotational and solenoidal components; (b) the fluid is barotropic;
 (c) a bath with the fluid undergoes vertical vibrations) the Navier-Stokes equation
 admits reduction to the modified Hamilton-Jacobi equation. The modification term is
 the Bohmian (quantum) potential. This reduction opens possibility to define
 a complex-valued function, named the wave function,
 which is a solution of the Schr\"{o}dinger equation.
 The solenoidal component being added to the momentum operator poses itself as a vector
 potential by analogy with the magnetic vector potential. The vector potential is
 represented by the solenoidal velocity multiplied by mass of the fluid element.
 Vortex tubes, rings, and balls along with the wave function guiding these objects
 are solutions of this equation. Motion of the vortex balls along
 the Bohmian trajectories gives a model of droplets moving on the fluid surface.
 The physical vacuum represents a peculiar superfluid medium. It consists of Bose particle-antiparticle pairs having nonzero masses and the angular momenta because of rotating about the center of the masses.
 Bundles of the vortex lines can transmit a torque from one rotating classical disk to other.\\
% A peculiar fluid is the superfluid physical vacuum.
% It contains Bose particle-antiparticle pairs. Vortex lines presented by
% electron-positron pairs are main torque objects.
% Bundles of the vortex lines can transmit a torque from one rotating disk
% to other unmoved disk.\\


\end{abstract}

\maketitle

\large

\section{\label{sec:level1}Introduction}

 Quantum mechanics beginning with its birth up to present time shows triumphal
 contribution to many technological areas of human activities.
 As for ontological basis, as ironic as it may sound, different interpretations of
 quantum mechanics, contradicting one another in some subtle points, are applied for
 describing one and the same physical experiment. The interpretations compete each
 with other for the sake of clearness of ontological understanding this discipline~\citep{QuoVadis}.
 Two well known interpretations of quantum mechanics are the Copenhagen interpretation and the interpretation based on the de Broglie-Bohm theory.
 Both interpretations have in their foundation the concept of the wave-particle dualism.
 The Copenhagen interpretation gives priority to the wave function and its collapse is conditioned by detecting a particle. The wave function describes only a probabilistic position of the particle in the physical space. In turn, De Broglie-Bohm theory says that the wave function guides the particle from its source up to a detector along a most optimal path. In this case the wave function plays a more active role.

 It is amazing, confirmation of the de Broglie-Bohm theory came from the area faraway from quantum mechanics. This is an experiment with droplets which bounce on a silicon oil surface at moving through an obstacle containing two slits~\citep{CouderForte:2006}. The authors have shown that in the far field an interference pattern arises after   passing of many droplets. It was found that the Faraday waves play a role of the guiding wave for the droplets bouncing on the silicon oil surface, a bath with which is subjected to small vertical vibrations.

 Where is truth? In order to answer on this question, it makes sense to begin from fundamentals.
 Let us begin from the most fundamental laws of classical physics.
 They are three Newton's laws first published in Mathematical Principles of Natural
 Philosophy in 1687~\citep{Motte:1846}. The Newton's laws read:
 (a) the first law postulates existence of inertial reference frames:
 an object that is at rest will stay at rest unless an external force acts upon it;
 an object that is in motion will not change its velocity unless an external force acts
 upon it.
 The inertia is a property of the bodies to resist to changing their velocity;
 (b) the second law states: the net force applied to a body with a mass $M$ is equal to
 the rate of change of its linear momentum in an inertial reference frame
\begin{equation}
\label{eq=1}
  {\vec F} = M{\vec a} = M{{d{\vec {\mathit v}}}\over{d\,t}};
\end{equation}
 (c) the third law states: for every action there is an equal and opposite reaction.

 Leonard Euler had generalized Newton's laws of motion on deformable bodies that are
 assumed as a continuum~\citep{Faber1995, Frisch:2008}. We rewrite the second Newton's law for
 the case of deformed medium. Let us imagine that a volume ${\Delta V}$ contains
 a fluid medium of the mass M. We divide Eq.~(\ref{eq=1}) by ${\Delta V}$ and determine
 the time-dependent mass density $\rho_{_{M}} = M/{\Delta V}$.
 In this case ${\vec F}$ is understood as the force per volume.
 Then the second law in this case takes a form:
\begin{equation}
\label{eq=2}
  {\vec F} = {{d\rho_{_{M}}{\vec {\mathit v}}}\over{d\,t}} =
  \rho_{_{M}}{{d{\vec {\mathit v}}}\over{d\,t}}+{\vec {\mathit v}}{{d\rho_{_{M}}}\over{d\,t}}.
\end{equation}
 The total derivatives in the right side can be written down through partial derivatives:
\begin{eqnarray}
\label{eq=3}
 {{d\rho_{_{M}}}\over{d\,t}}   &=& {{\partial\rho_{_{M}}}\over{\partial\,t}}
  + ({\vec {\mathit v}}\,\nabla)\rho_{_{M}},\\
 {{d{\vec {\mathit v}}}\over{d\,t}}\; &=&\; {{\partial{\vec {\mathit v}}}\over{\partial\,t}}
  \;+\; ({\vec {\mathit v}}\,\nabla){\vec {\mathit v}}.
\label{eq=4}
\end{eqnarray}
 Eq.~(\ref{eq=3}) equated to zero is seen to be the continuity equation.
 As for Eq.~(\ref{eq=4}) we may rewrite the rightmost term 
 in detail
\begin{equation}
\label{eq=5}
 ({\vec {\mathit v}}\,\nabla){\vec {\mathit v}} = \nabla\;{{{\mathit v}^2}\over{2}}
 -[{\vec {\mathit v}}\times[\nabla\times{\vec {\mathit v}}]].
\end{equation}
 As follows from this formula the first term, multiplied by the mass, is gradient of
 the kinetic energy. It is a force applied to the fluid element for its shifting on the
 unit of length,~$\delta s$. The second term is acceleration of the fluid element
 directed perpendicularly to the velocity. Let the fluid element move along some curve in
 3D space. Tangent to the curve in each point refers to orientation of the body motion.
 In turn, the vector ${\vec\omega}=[\nabla\times{\vec {\mathit v}}]$ is orientated perpendicularly
 to the plane, where an arbitrarily small segment of the curve lies.
 This vector characterizes a quantitative measure of the
 vortex motion and it is called {\it vorticity}. 
 Vector product $[{\vec {\mathit v}}\times{\vec\omega}]$ is perpendicular to the both vectors ${\vec {\mathit v}}$  and ${\vec\omega}$.
 It represents the Coriolis acceleration of the body
 under rotating it around the vector  ${\vec\omega}$.
 In turn, the term $\rho_{_{M}}[{\vec {\mathit v}}\times{\vec\omega}]$ represents the Coriolis force acting on the fluid element in the volume $\Delta V$. 

 The term (\ref{eq=5}) entering in the Navier-Stokes equation is responsible for emergence of vortex structures. The vortex tubes, rings, and balls are considered briefly in Sec.~2 The latter objects are good models of the droplets moving on the fluid surface. In Sec.~3 we transform the Navier-Stokes to the Schr\"{o}dinger-like equation. The transformation is achieved due to introduction of a material dependent parameter which substitutes the Plank constant.
 Solutions of the equation disclose interference patterns arising from the motion of the vortex balls along optimal paths, the Bohmian trajectories, through an obstacle containing slits. The Schr\"{o}dinger equation describing flow of a superfluid medium known as the physical vacuum is considered in Sec.~4. Here we deal with vortex lines which arise due to rotation of the virtual electron-positron pairs. The lines can self-organize into twisted vortex bundles in a rotating cylindrical container.
 Kinetic energy of these vortex bundles is sufficient to begin to rotate a top disk covering the container.
 Concluding remarks are given in Sec.~5. 

\section{\label{sec:level2}The Navier-Stokes equation and motion of vortices}

 Taking into account the continuity equation let us rewrite Eq.~(\ref{eq=2}) by omitting
 the rightmost term and specify forces which should be in this equation:
\begin{equation}
\label{eq=6}
 \rho_{_{M}}\Biggl(
 {{\partial {\vec {\mathit v}}}\over{\partial\,t}}
 + ({\vec {\mathit v}}\cdot\nabla){\vec {\mathit v}}
       \Biggr)
  = -\nabla P + \mu\,\nabla^{\,2}{\vec {\mathit v}} + {{{\vec F}}\over{\Delta V}}.
\end{equation}
 The following set of forces is presented in this equation: (i)~the first term is
 the pressure gradient. It takes place when there is a difference in pressure across a
 surface; (ii)~the second term represents the viscosity of a Newtonian fluid
 (here $\mu$ is the dynamic viscosity, its units are N$\cdot$s/m$^2$ = kg/(m$\cdot$s));
 (iii)~${\vec F}/{\Delta V}$ is a body force per unit volume acting on the fluid.
 So, we wrote down the Navier-Stokes equation for the viscous incompressible Newtonian
 fluid~\citep{LandauLifshitz1987}.
 The term $({\vec {\mathit v}}\cdot\nabla){\vec {\mathit v}}$ in Eq.~(\ref{eq=6})
 can be rewritten in detail as shown in Eq.~(\ref{eq=5}).
 The vector ${\vec\omega}=[\nabla\times{\vec {\mathit v}}]$ in Eq.~(\ref{eq=5}), named
 {\it vorticity}, underlies the vortex formation. Let us consider the simplest examples.

\subsection{\label{subsec:level2A}Helmholtz vortices}

 Let the force ${\vec F}$ be conservative. Then by applying to Eq.~(\ref{eq=6}) the curl
 we obtain right away the equation for the vorticity:
\begin{equation}
\label{eq=7}
 {{\partial\,{\vec\omega}}\over{\partial\,t}}
 + ({\vec\omega}\cdot\nabla){\vec {\mathit v}}
 = \nu\nabla^{\,2}{\vec\omega}.
\end{equation}
 Here $\nu=\mu/\rho_{_{M}}$ is the kinematic viscosity. Its dimension is [m$^2$/s].
 It corresponds to the diffusion coefficient. For that reason the energy stored in the
 vortex will dissipate in thermal energy. As a result the vortex, with the lapse of time,
 will disappear. For supporting the vortex activity, the energy has to be supplied permanently.
 Trivial solution of Eq.~(\ref{eq=7}) is seen to be  ${\vec\omega}=0$.

 We suppose that the fluid is ideal, barotropic, and the mass forces are conservative~\citep{KunduCohen:2002}. 
 With omitted the right term, i.e., at $\nu=0$,
 the Helmholtz theorem says: (i)~if fluid particles
 form in any moment of the time a vortex line then the same particles
 form the vortex line both in the past and in the future; (ii)~ensemble
 of the vortex lines traced through a closed contour forms a vortex tube.
 Intensity of the vortex tube is constant along its length and does not
 change in time. Intensity of the vortex is circulation of the velocity
 around the contour encompassing the vortex. From above said follows that
 the vortex tube (a)~either goes to infinity by both endings; (b)~or
 these endings lean on walls of bath containing the fluid; (c)~or these
 endings are closed to each on other forming a vortex ring.

 As for the vortex solutions of Eq.~(\ref{eq=7}) it is convenient to proceed to
 cylindrical or toroidal frame of reference depending on a task which is
 to be considered. In the first case the solutions are vortex tubes
 oriented along the axis $z$. In the second case they are vortex rings
 lying in the plane $(x,y)$. Consider the both cases.

\subsubsection{\label{subsubsec:level2A1}Vortex tube}

\begin{figure}
  \centering
  \begin{picture}(200,330)(75,35)
  \includegraphics{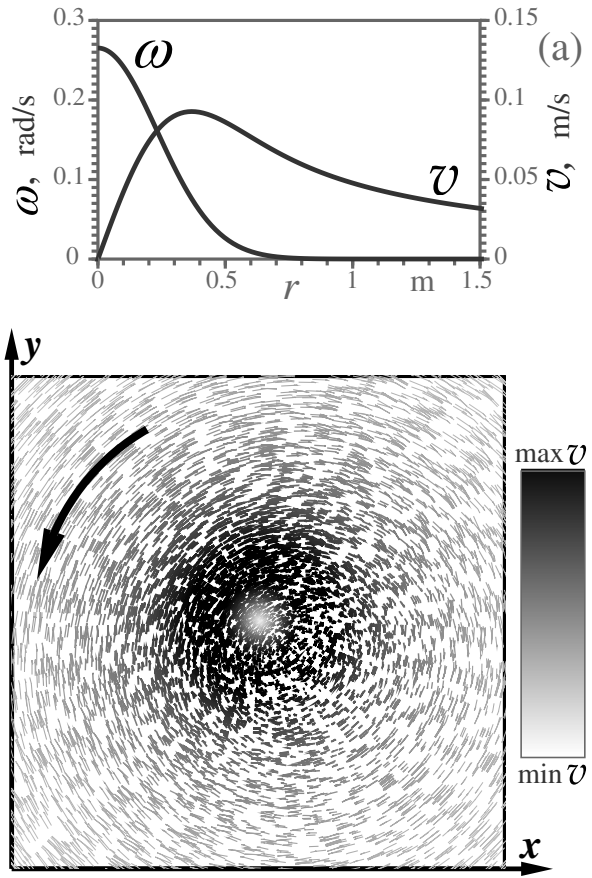}
  \end{picture}
  \caption{
 Vortex tube oriented along the axis $z$: (a) Vorticity $\omega$ and velocity $u$ of
 a flow around the center as functions of remoteness from this center. The velocity $u$
 vanishes in the center and tends to zero on infinity. (b) 
 Cross-section of the vortex.
 Values of the velocity $u$ are shown in  grey ranging from light grey (min $u$) to dark grey (max $u$).  
 Density of the pixels represents magnitude of the vorticity $\omega$ }
  \label{fig=1}
\end{figure}

 Let the vortex tube be oriented along the axis $z$ and its axis is aligned with
 the origin $(x,y) = (0,0)$. Eq.~(\ref{eq=7}) rewritten in cross-section of the vortex
 flow reads
\begin{equation}
\label{eq=8}
 {{\partial\omega}\over{\partial\,t}} =
 \nu\Biggl(
    {{\partial^{\,2}\omega}\over{\partial r^2}}
  + {{1}\over{r}}{{\partial\omega}\over{\partial r}}
    \Biggr).
\end{equation}
 Here we do not write a sign of vector on the top of $\omega$ since $\omega$ is oriented
 strictly along the axis $z$. Solution of this equation is as follows
\begin{equation}
\label{eq=9}
  \omega = {{\mit\Gamma}\over{4\pi\nu\,t}}
  \exp\Biggl\{
    \,-{{r^{\,2}}\over{4\nu\,t}}\,
      \Biggr\}.
\end{equation}
 Here ${\mit\Gamma}$ is the integration constant having dimension [m$^2$/s].
 Components of the velocity ${\vec {\mathit v}}$ are lying in the plane $(x,y)$ along tangents of
 circles. We have the Lamb-Oseen solution
\begin{equation}
\label{eq=10}
   {\mathit v}(r,t) = {{1}\over{r}}
    \int\limits_{0}^{r} \omega r'dr' 
  = {{\mit\Gamma}\over{2\pi\,r}}\biggr(
    1 - e^{-r^{\,2}/4\nu\,t}
                                \biggl).
\end{equation}
 Fig.~\ref{fig=1}(a) shows the vorticity $\omega$ and the velocity $u$
 as functions of distance from the vortex center.
 Because of $\nu > 0$ the vortex decays with time.
 Qualitative view of the vortex in its cross-section is shown in Fig.~\ref{fig=1}(b).
 Values of the velocity $u$ are shown in grey from light grey (minimal velocities)
 to dark grey (maximal ones). One can see that the cross-section of the vortex gives
 a good illustration of a hurricane shown from the top. In the center of the vortex
 a so-called eye of the hurricane is well viewed. Small values of the velocity
 in the eye of the hurricane are marked by light grey. Observe that in the region
 of the eye a wind is really very weak, especially near the center. 
 This is in stark contrast to conditions in the region of the eyewall,
 where the strongest winds exist (it is shown in Fig.~\ref{fig=1}(b) by a dark grey annular region encompassing
 the light grey region of the eye).

\subsubsection{\label{subsubsec:level2A2}Vortex ring}

\begin{figure}
  \centering
  \begin{picture}(200,150)(0,25)
  \includegraphics{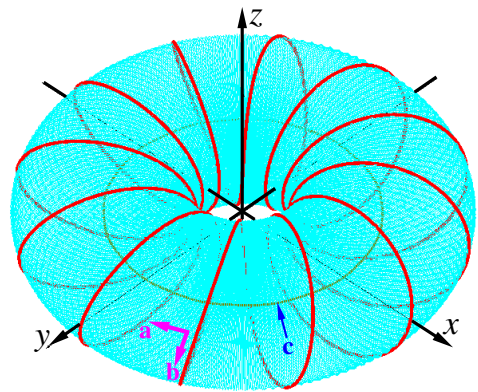}
  \end{picture}
  \caption{Helicoidal vortex ring: 
   $r_{0}=2$, $r_{1}=3$, $\omega_{2} = 12\omega_{1}$, and $\phi_{1}=\phi_{2}=0$.
  }
  \label{fig=2}
\end{figure}

 If we roll up the vortex tube in a ring and glue together its opposite ends we obtain
 a vortex ring. A result of such an operation put into the $(x,y)$ plane is shown in
 Fig.~\ref{fig=2}. Position of points on the helicoidal vortex ring~\citep{Sonin:2012}
 in the Cartesian coordinate system is given by
\begin{eqnarray}
\nonumber
 x &=& (r_{1} + r_{0}\cos(\omega_{2}t +\phi_{2}))\cos(\omega_{1}t +\phi_{1}),~~~~ \\
\label{eq=11}
 y &=& (r_{1} + r_{0}\cos(\omega_{2}t +\phi_{2}))\sin(\omega_{1}t +\phi_{1}),~~~~ \\
\nonumber
 z &=& r_{0}\sin(\omega_{2}t +\phi_{2}) .~~~~
\end{eqnarray}
 Here $r_{1}$ represents the distance from the center of the tube (pointed in the figure by arrow
 {\it c}) to the center of the torus located in the origin of  coordinates $(x,y,z)$.
 And $r_{0}$ is the radius of the tube. 
 A body of the tube, for the sake of visualisation, is dashed by light grey curves. 
 Eq.~(\ref{eq=11}) parametrized by $t$
  gives a helicoidal vortex ring shown in Fig.~\ref{fig=2}.
 Parameters $\omega_{1}$ and $\omega_{2}$ are frequencies of rotation along the arrow {a} about
 the centre of the torus (about the axis $z$) and rotation along the arrow b about the centre of
 the tube (about the axis pointed by arrow c), respectively. Phases $\phi_{1}$ and $\phi_{2}$ 
 have uncertain quantities ranging from 0 to $2\pi$. By choosing the phases within this interval
 with small increment, we may fill the torus by the helicoidal vortices everywhere densely.

 As was mentioned above, the vorticity is maximal along the centre of the tube. Whereas the velocity of rotation about this centre in the vicinity of it is minimal. A stream on the centre will go in parallel to this central axis.
 However, the velocity grows at increasing distance from the centre.
 After reaching some maximal value that depends on magnitude of the viscosity, see Eq.~\ref{eq=10}, the velocity begins to decrease. So the vortex ring has a finite size. Further we shall return to the helicoidal vortex ring for the case when the radius $r_{1}$ will tend to zero.

\subsubsection{\label{subsec:level2A3}Vortex ball}

\begin{figure}
  \centering
  \begin{picture}(200,160)(0,10)
  \includegraphics{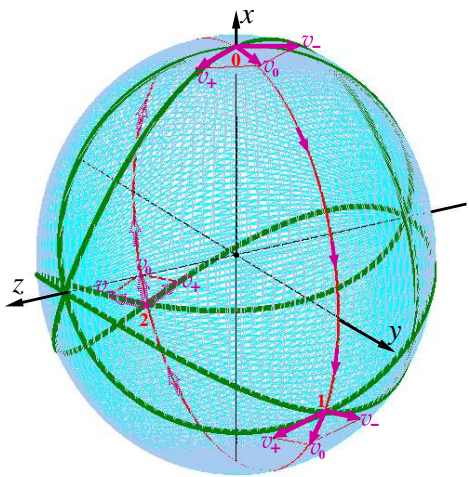}
  \end{picture}
  \caption{ Helicoidal vortex ring convoluted onto the vortex ball: 
  $r_0=4$, $r_1=0.01 \ll 1$, $\omega_2=3\omega_1$, $\phi_1= \phi_2=0$.   
   The radius $r_{0}$ represents a mean radius of the ball,
   where the velocity $\mathit v_{\,0}$ reaches a maximal value.
   The ratio $\omega_2/\omega_1 = 3$ was chosen with the aim in order not to overload the picture. 
  }
  \label{fig=3}
\end{figure}

 Here we shall represent a vortex ball that can simulate the droplet. Let the radius $r_1$ in Eq.~(\ref{eq=11})  tends to zero. 
The helicoidal vortex ring in this case will transform into a vortex ring enveloping a spherical ball. 
The vortex ring for the case $r_0=4$, $r_1 \approx 0$, $\omega_2=3\omega_1$, and $\phi_1=\phi_2=0$ drawn by thick  curve colored in dark green is shown in Fig.~\ref{fig=3}. 
Motion of a small particle (SP) along the vortex ring takes place with a velocity
\begin{widetext}
\begin{equation}
  {\vec {\mathit v}}_R = 
   \left(
   \matrix{
     {\mathit v}_{_R,x}= -r_0\omega_2\sin(\omega_2t+\phi_2)\cos(\omega_1t+\phi_1)
              -r_0\omega_1\cos(\omega_2t+\phi_2)\sin(\omega_1t+\phi_1)
     \cr
     {\mathit v}_{_R.y}= -r_0\omega_2\sin(\omega_2t+\phi_2)\sin(\omega_1t+\phi_1)
              +r_0\omega_1\cos(\omega_2t+\phi_2)\cos(\omega_1t+\phi_1) 
     \cr
     {\mathit v}_{_R,z}=~~r_0\omega_2\cos(\omega_2t+\phi_2)\hspace{242pt} \cr
          }
   \right).
\label{eq=12}
\end{equation}
\end{widetext}
 The velocity of SP at the initial time is ${\mathit v}_{_R,x}= 0$, ${\mathit v}_{_R,y}= r_0\omega_1$, ${\mathit v}_{_R,z}= r_0\omega_2 = 3r_0\omega_1$  (the initial point $(x, y, z) = (4, 0, 0)$ is on the top of the ball). We designate this velocity as  ${\vec {\mathit v}}_{+}$. Through $t = \pi/\omega_1$ the SP returns to the top position. The velocity in this case is ${\mathit v}_{_R,x} = 0$, ${\mathit v}_{_R,y} = r_0\omega_1$, ${\mathit v}_{_R,z}= -r_0\omega_2 = -3r_0\omega_1$. We designate this velocity as  ${\vec {\mathit v}}_{-}$. Sum of the two opposite velocities, ${\vec {\mathit v}}_{+}$  and ${\vec {\mathit v}}_{-}$ , gives the velocity  ${\vec {\mathit v}}_{0}=(0,2r_0\omega_1,0)$. During $t = (1+3k)\pi/3\omega_1$ and $t = (2+3k)\pi/3\omega_1$ ($k=1,2,\cdots$) SP travels through the positions 1 and 2 both in the forward and in backward directions, respectively. In these points the velocities  ${\vec {\mathit v}}_{+}$  and ${\vec {\mathit v}}_{-}$  give the resulting velocity ${\vec {\mathit v}}_{0}$  directed along the circle lying in the plane $(x, y)$. 

\begin{figure}
  \centering
  \begin{picture}(200,160)(-15,5)
  \includegraphics{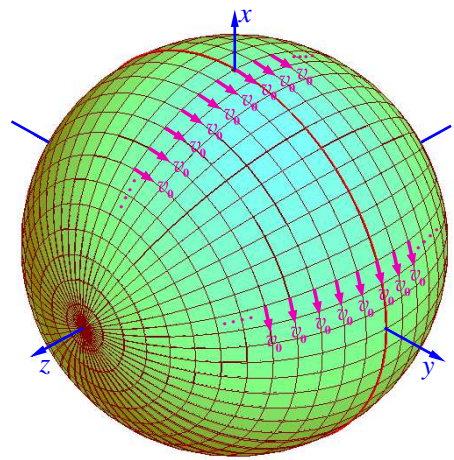}
  \end{picture}
  \caption{ The vortex ball rotating about axis $z$ 
   with the maximal velocity $\mathit v_{\,0}$ that is reached on the surface of the ball.
     }
  \label{fig=3c}
\end{figure}

By adding the vortex rings with other phases $\phi_1$ and $\phi_2$ ranging from 0 to $2\pi$ we fill the ball everywhere densely by these rings. The resulting velocities ${\vec {\mathit v}}_{\,0}$ in some points of the ball 
will lie on circles centered on the axis $z$.
We see a ball that rolls along the axis $y$, Fig.~\ref{fig=3c}. 
The spherical harmonics are perfect modes in this case~\citep{DorboloEtAl2008}.

\section{\label{sec:level3}Transformation of the Navier-Stokes equation}

 Before we shall begin to subject the Navier-Stokes equation to transformations
 it would be instructive to recall works touching upon analogies between hydrodynamics and quantum mechanics.
 First quantum theory in a hydrodynamic representation  was formulated by Erwin Madelung~\citep{Madelung1926}.
 Remarkably that Madelung's equations exhibit a close relation
 of the Bohmian mechanics~\citep{GuantesEtAl:2004, OriolsMompart2012, Wyatt2005} and hydrodynamics.
 It gives the reason to hypothesize that quantum medium behaves like a fluid with irregular fluctuations~\citep{BohmVigier1952}. 

 One more impressive example comes from the fluid mechanics.
 It is a demonstration of quantum-like behavior of droplets bouncing on fluid surface~\citep{CouderForte:2006, CouderEtAll:2005, ProtierEtAl:2006, EddiEtAll:2011}.
 After the articles appeared in the press, attempts at an explanation of this experiment were undertaken
 with the point of view of quantum mechanics~\citep{DorboloEtAl2008, MolacekEndBush2013, Sbitnev2013b, Brady2013, BradyAnderson2014, HarrisBush2014, OzaEtAl2014, Grossing2009, Grossing2010, GrossingEtAl2011, Grossing2013}.
 Since we know that presence of the quantum potential~\citep{Bohm1952, BohmHiley1982} is a sign of the nonlocal interaction
 this summary gives a perspective to find an analogue of the quantum potential for the case of the classical fluid mechanical system.
 One may suppose from quantum mechanical point of view that behavior of an incompressible liquid can be described by
 a Schr\"{o}dinger-like equation. However we need to replace in this case the Planck constant by
 other parameter~\citep{Sbitnev2013b, BradyAnderson2014} exceeding the Planck constant at many orders.

 Two equations, describing flow of an incompressible fluid~\citep{LandauLifshitz1987}, are the Navier–Stokes equation~(\ref{eq=6}) and the mass conservation equation; it is Eq.~(\ref{eq=3}) equated to zero. The mass density $\rho_{_{M}}$ we replace further by a probability density $\rho$ according to the following formula
\begin{equation}
\label{eq=13}
 \rho_{_{M}} = {{M}\over{\Delta V}} = {{mN}\over{\Delta V}} = m\rho.
\end{equation}
 Here the mass $M$  is a product of an elementary mass $m$ by the number of these masses,~$N$, filling the volume $\Delta V$. Then the mass density $\rho_{_{M}}$ is defined as a product of the elementary mass $m$ by the density of quasi-particles 
$\rho=N/\Delta V$. The quasi-particle is a droplet-like inhomogeneity, which moves with the local stream velocity of the equivalent fluid. We assume that each quasi-particle moves like a Brownian particle with the diffusion coefficient inversely proportional to $m$ and subjected to the kinematic viscosity $\nu = \mu/\rho_{_{M}}$. So that, $\rho$ is the probability density of finding the quasi-particle within the volume $\Delta V$ and obeys the conservation law
\begin{equation}
\label{eq=14}
 {{\partial\,\rho}\over{\partial\,t}}
 + ({\vec {\mathit v}}\cdot\nabla)\rho = 0.
\end{equation}

 We accept the following assumptions: (a)~the velocity consists of two components -  irrotational and solenoidal~\citep{KunduCohen:2002} that relate to vortex-free and vortex motions, respectively. The basis for the latter is the Helmholtz theorem; (b)~ we admit that the density is a function of the pressure, i.e., the fluid is barotropic. The Madelung's nonlocal quantum pressure~\citep{Madelung1926} has a deep relation to the osmotic pressure in vacuum, as follows from Nelson's work~\citep{Nelson1966}. Then the Navier-Stokes equation can be reduced to the Schr\"{o}dinger equation if we assume that the pressure $P$ and the Bohmian quantum potential~\citep{Bohm1952} are compatible; (c)~in order to neutralize the viscosity we shall shake a bath with the fluid vertically with some frequency and amplitude, so that the Faraday oscillations will stay subcritical~\citep{EddiEtAll:2011}. 

 {\bf (a)} The irrotational and solenoidal components submit to the following equations
\begin{equation}
 \left\{
 \matrix{
        {\vec {\mathit v}} = {\vec {\mathit v}_{S}} + {\vec {\mathit v}_{R}} = {\displaystyle{{1}\over{m}}}\nabla S +{\vec {\mathit v}_{R}}, \cr\cr
        (\nabla\cdot {\vec {\mathit v}_{R}}) = 0, ~~~ [\nabla\times{\vec {\mathit v}_{R}}] ={\vec\omega}. \cr
        }
  \right.
\label{eq=15}
\end{equation}\\
 Subscripts $S$ and $R$ hint to scalar and vector (rotational) potentials underlying emergence of these two components of the velocity. The scalar field is represented by the scalar function $S$ –- action in classical mechanics.
 Both velocities are perpendicular to each other. Now we may define the momentum and the kinetic energy of the quasi-particle
\begin{eqnarray}
\label{eq=16}
 {\vec p} &=& m{\vec {\mathit v}} = \nabla S + m{\vec {\mathit v}_{R}}, \\
  m{{{\mathit v}^2}\over{2}} &=& {{1}\over{2m}}(\nabla S)^2 + m {{{\mathit v}_{R}^2}\over{2}}.
\label{eq=17} 
\end{eqnarray}
 Now we may rewrite Navier-Stokes equation~(\ref{eq=6}) in the following form
\begin{widetext}
\begin{equation}
  {{\partial}\over{\partial\,t}}(\nabla S + m{\vec {\mathit v}_{R}}) +
  \underbrace{
     {{1}\over{2m}}\Bigl((\nabla S)^2 + m^2{\mathit v}_{R}^2\Bigr)
      + \Bigl[{\vec\omega}\times(\nabla S + m{\vec {\mathit v}_{R}})\Bigr]
             }_{(a)}
 = -{{1}\over{\rho}}\nabla P - \nabla U + \nu\nabla^{\,2}(\nabla S + m{\vec {\mathit v}_{R}}).
\label{eq=18}
\end{equation}
\end{widetext}
 Note that the term embraced by the curly bracket (a) is the term $m({\vec {\mathit v}}\cdot\nabla){\vec {\mathit v}}$
 in Eq.~(\ref{eq=6}).

 {\bf (b)} The quantum-like behavior of droplets shown in~\citep{CouderForte:2006} hints that the Navier-Stokes equation under some modes can be reduced to the Schr\"{o}dinger equation~\citep{Sbitnev2013b, BradyAnderson2014}. 
Consider in this connection the first term from the right side of Eq.~(\ref{eq=18}) in detail. More definitely, consider the pressure $P$ which we shall represent consisting of two parts, $P_1$ and $P_2$. 
We begin from the Fick$^{\prime}$s law~\citep{Grossing2009, Grossing2010}.
The law says that the diffusion flux, ${\vec J}$, is proportional to the negative value of the density gradient  ${\vec J}=-D\nabla\rho$, where $D=\eta_{\sigma}/(2m)$ is the diffusion coefficient~\citep{Nelson1966}. The parameter $\eta_{\sigma}$ is the parameter replacing the Planck constant for the case of the fluid flows~\citep{Sbitnev2013b}. Since the term  $\eta_{\sigma}\nabla{\vec J}$  has dimension of the pressure, we define $P_1$ as the pressure having diffusion nature
\begin{equation}
\label{eq=19}
  P_1 = {{\eta_{\sigma}}\over{2}}\nabla{\vec J} = -{{\eta_{\sigma}^2}\over{4m}}\nabla^{\,2}\rho.
\end{equation}
 Kinetic energy of the diffusion flux is $(m/2)(J/\rho)^2$. It means, that there exists one more pressure as the average momentum transfer per unit area per unit time
\begin{equation}
\label{eq=20}
 P_2 = \rho{{m}\over{2}}\Biggl(
                         {{J}\over{\rho}}
                        \Biggr)^2
 = {{\eta_{\sigma}^2}\over{8m}}{{(\nabla\rho)^2}\over{\rho}}.
\end{equation}

 Observe that sum of the two pressures, $P_1+P_2$, divided by $\rho$ gives a term
\begin{equation}
\label{eq=21}
 Q = -{{\eta_{\sigma}^2}\over{2m}}
 \Biggl[
   {{\nabla^{\,2}\rho}\over{2\rho}} - \Biggl(
                                            {{\nabla\rho}\over{2\rho}}
                                      \Biggr)^2
 \Biggr]
\end{equation}
 having the same form as the quantum potential.
 E. Nelson has shown that this pressure has an osmotic nature~\citep{Nelson1966}. 
 It can be interpreted as follows: a semipermeable membrane where the osmotic pressure manifests itself is an instant which divides the past and the future (that is, the 3D brane of our being is the semipermeable membrane).
 Now we may rewrite the first term from the right side in Eq.~(\ref{eq=18}) as follows
\begin{equation}
\label{eq=22}
  {{\nabla P}\over{\rho}} = {{1}\over{\rho}}\nabla\Biggl(
                                                  \rho{{P}\over{\rho}}
                                                  \Biggr) =
  {\nabla Q} + {{\nabla\rho}\over{\rho}}Q.
\end{equation}
 Gr{\"o}ssing noticed that the term $\nabla Q$, the gradient
 of the quantum potential, describes a completely thermalized fluctuating force field~\citep{Grossing2009, Grossing2010}. 
 Here the fluctuating force is expressed via the gradient of the pressure divided by the density distribution
 of particles chaotically moving.
 
 The term $\nabla Q$ in Eq.~(\ref{eq=22}) fits for further transformation of the Navier-Stokes equation to the Schr\"{o}din\-ger-like equation. 
 But the second term from the right side of this equation is superfluous. For that reason we need to admit that the fluid should be incompressible, $\nabla\rho=0$.

 The second term from right side of Eq.~(\ref{eq=18}) is a gradient of the potential energy $U$, which represents negative value of the conservative force acting to the quasi-particle. 
 
 {\bf (c)} The last term from the right side of Eq.~(\ref{eq=18}) describes kinetic losses in the fluid because of the viscosity. To compensate decay of the waves in the fluid it is proposed to use the gravitational force by shaking the fluid periodically in vertical direction~\citep{CouderForte:2006, EddiEtAll:2011}. The frequency $\omega_{_{F}}$ of the periodic shaking in this case should satisfy to the following condition
\begin{equation}
\label{eq=23}
 \eta_{\sigma}\omega_{_{F}} = {\nu m(\nabla{\vec {\mathit v}})}= {{\mu}\over{\rho}}(\nabla{\vec {\mathit v}}).
\end{equation}
 Here $\mu = \nu m\rho$ is the dynamical viscosity of the fluid. The viscosity poses by itself as a resistance to motion of the fluid. Observe that measure of the viscosity is determined by the force required to shear the fluid. The shaking of the container should be weak enough in order to initiate the Faraday waves slightly below the critical threshold. Due to this trick, effect of the viscosity may be neutralized~\citep{EddiEtAll:2011}. Further, for that reason we shall not take into account the term  $\nu\nabla^{\,2}\nabla S$.

 Let us multiply Eq.~(\ref{eq=18}) from the left by ${\vec {\mathit v}}_S$  and take that $({\vec {\mathit v}}_S\cdot{\vec {\mathit v}}_R)=0$  and $({\vec {\mathit v}}_S\cdot{\vec\omega})=0$. Next, by integrating the equation over the volume of the fluid we obtain the following modified by the quantum potential, $Q$, Hamilton-Jacobi equation
\begin{eqnarray}
\nonumber
 && {{\partial}\over{\partial\,t}}S + {{1}\over{2m}}(\nabla S)^2 + {{m}\over{2}}{\mathit v}_R^2 + U\\
 &&-{{\eta_{\sigma}^2}\over{2m}}
 \Biggl[
   {{\nabla^{\,2}\rho}\over{2\rho}} - \Biggl(
                                            {{\nabla\rho}\over{2\rho}}
                                      \Biggr)^2\,
 \Biggr] = C.
\label{eq=24}
\end{eqnarray}
 In this equation $C$ is an integration constant. Here we used $\nabla Q$  instead of $\nabla P/\rho$  as follows from Eq.~(\ref{eq=22}), where $Q$ is given in Eq.~(\ref{eq=21}). We see that Eq.~(\ref{eq=24}) contains a term which represents energy of the vortex. Also we can see that the vortex in Eq.~(\ref{eq=7}) is replenished by the kinetic energy coming from the scalar field $S$, namely via the term  $({\vec\omega}\cdot\nabla){\vec {\mathit v}}$. Solutions of these two equations,  Eqs~(\ref{eq=7}) and~(\ref{eq=24}),  representing the vortex and scalar fields, depend on each other.
\begin{figure}
  \centering
  \begin{picture}(200,240)(30,0)
  \includegraphics{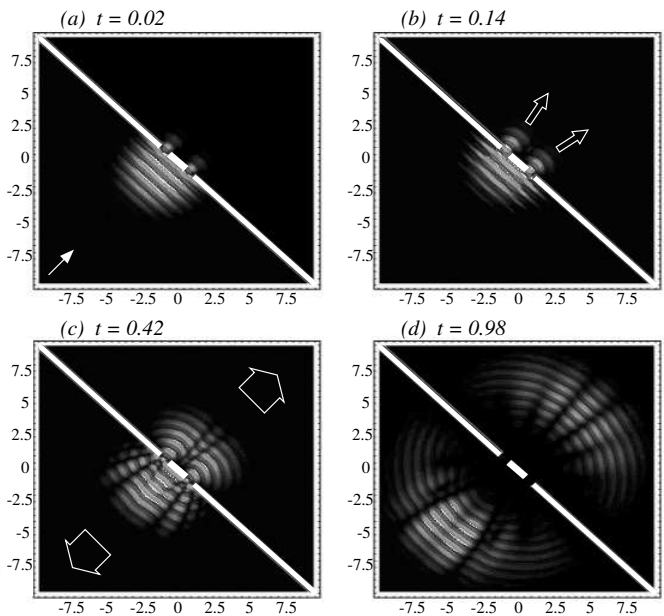}
  \end{picture}
  \caption{ Scattering of a Gaussian soliton-like wavepacket on a barrier containing two slits~\citep{Sbitnev2009, Sbitnev2013b}.
%[V. I. Sbitnev, \urlprefix\url{http://arxiv.org/abs/1307.6920]}. 
Time here is in seconds and scale of the area is given in centimeters. Velocity of the wave is about 150 mm/s. It corresponds to the phase velocity of the Faraday waves stated in~\citep{CouderForte:2006}.
%[Y. Couder and E. Fort, Phys. Rev. Lett., 97, 1541010 (2006)].   
  }
  \label{fig=4}
\end{figure}

Because of presence of the quantum potential in Eq.~(\ref{eq=24}) this equation is called the quantum Hamilton-Jacobi equation. Together with the continuity equation~(\ref{eq=14}) it can be extracted from the following wave Schr\"{o}dinger-like equation
\begin{equation}
 {\bf i}\eta_{\sigma}{{\partial\Psi}\over{\partial\,t}} =
 {{1}\over{2m}}(-{\bf i}\eta_{\sigma}\nabla+m{\vec {\mathit v}}_R)^2\Psi +U\Psi - C\Psi.
\label{eq=25}
\end{equation}
 The kinetic momentum operator $(-{\bf i}\eta_{\sigma}\nabla+m{\vec {\mathit v}}_R)$  contains the term $m{\vec {\mathit v}}_R$  
which describes a contribution of the rotation field conditioned by the vortical motion. 
This term is analogous to the vector potential multiplied by the ratio of the charge to the light speed 
which is represented in quantum electrodynamics~\citep{Martins2012}.
Existence of the vector rotation field is supported by the Helmholtz theorem. By substituting into Eq.~(\ref{eq=25}) the wave function $\Psi$ represented in a polar form
\begin{equation}
\label{eq=26}
  \Psi = \sqrt{\rho}\cdot\exp\{{\bf i}S/\eta_{\sigma}\}
\end{equation}
 and separating on real and imaginary parts we obtain Eqs.~(\ref{eq=24}) and~(\ref{eq=14}).

 A solution of the linear Schr\"{o}dinger equation with the potential simulating a barrier with two slits discloses interference patterns from them both forward and backward. Scattering of a Gaussian soliton-like wave on the slits produces two waves - reflected and transmitted, Fig.~\ref{fig=4}. Both waves evolve against the background of the subcritical Faraday waves. As for the Faraday waves, Eq.~(\ref{eq=25}) under conditions stated in~\citep{MilesHenderson1990, Miles1993}
  can be reduced to the Miles-Henderson wave equation that describes their different modes.

\subsection{\label{subsec:level3A}The material dependent parameter $\eta_{\sigma}$}

 The vortex ball can be viewed as a droplet, when moving it on a fluid surface. In this case instead of the viscosity an important parameter is the surface tension. The surface tension of a liquid is the maximum restoring force that provides spherical form to the droplet. Actually the best way to think of this physically is that it is the energy cost of forming a liquid-air interface. 

 The surface tension $\sigma$ and the viscosity $\mu$ are not directly related. The units of the surface tension are [N/m]. Whereas the units of the viscosity are [N$\cdot$s/m$^2$]. However, Frenkel's theory of sintering~\citep{Frenkel1945} takes into account relationship of these parameters through the following differential dependence~\citep{RisticAndMilosevic2006}: 
\begin{equation}
\label{eq=27}
  {{d\,r}\over{d\,t}} =-{{3}\over{4}}{{\sigma}\over{\mu}}.
\end{equation}
 It describes change of the drop radius, $r$, with time. 
 Extent of the sintering is defined by~\citep{MillerAndKalmanovitch}
\begin{equation}   
  r^2 = {{3}\over{2}}{{\sigma}\over{\mu}}r\,t.
\label{eq=28}
\end{equation}
 Here $r^2$ is a contact area for a given time $t$. 
Let us now substitute this result in Eq.~(\ref{eq=23})
\begin{equation}
  \eta_{\sigma}\omega_{_{F}} %= {{\mu}\over{\rho}}(\nabla{\vec {\mathit v}})
 = {{3}\over{2}}{{\sigma}\over{r^2\rho}}r\,t(\nabla{\vec {\mathit v}})
 = {{1}\over{2}}\sigma S\cdot t (\nabla{\vec {\mathit v}}).
\label{eq=29}
\end{equation}
 Here we took into account, that the volume of the droplet is $\Delta V = (4/3)\pi r^3 =1/\rho$ and the area $S$ of the sphere is $4\pi r^2$. Observe that $t(\nabla{\vec {\mathit v}})$ is a dimensionless parameter. It gives a contribution only on stages of the periodic bouncing of the droplet~\citep{CouderEtAll:2005}. The periodic shaking prevents the process of the merging. Therefore we adopt that  $(t/2\pi)(\nabla{\vec {\mathit v}})S =\Delta S$ is the contact area of the droplet with the fluid through a thin air film. As a result we get
\begin{equation}
  \eta_{\sigma} = \sigma{\Delta S}{{2\pi}\over{2\omega_{_{F}}}} = \sigma{\Delta S}{{T_0}\over{2}}.
\label{eq=30}
\end{equation}
 Its value is about $10^{-11}$ J$\cdot$s~\citep{Sbitnev2013b} 
for the case of the silicon oil~\citep{CouderForte:2006}. Here $T_0/2$ is a half-period of the Faraday oscillations.
\begin{figure}
  \centering
  \begin{picture}(200,80)(15,20)
  \includegraphics{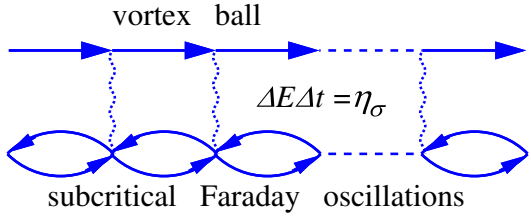}
  \end{picture}
  \caption{ 
 Feynman-like diagram representing the motion of the vortex ball that rolls along the fluid surface. Exchange by energy via a generated surface wave occurs at each bounce. Here relation ${\Delta E}{\Delta t} = \eta_{\sigma}$ simulates the well-known quantum-mechanical relation  ${\Delta E}{\Delta t} = \hbar$.   
  }
  \label{fig=5}
\end{figure}

\subsection{\label{subsec:level3B}The Bohmian trajectories and interference phenomena}

 In the example of the vortex ball given in Fig.~{\ref{fig=3} the droplet will revolve on 120$^{\rm o}$ after each bouncing. As the droplet moves on the fluid surface, it induces Faraday waves on this surface, which, in turn, effect on motion of the droplet~\citep{CouderEtAll:2005, EddiEtAll:2011}. Following to Feynman's ideas~\citep{Feynman1948, FeynmanHibbs1965} the moving droplet induces in the fluid waves accompanying the droplet, 
Fig.~\ref{fig=5}. This follows, in particular, from the third Newton's law - when an object exerts force on the surface, the surface will push back that object with equal force in the opposite direction.
\begin{figure}
  \centering
  \begin{picture}(200,270)(20,0)
  \includegraphics{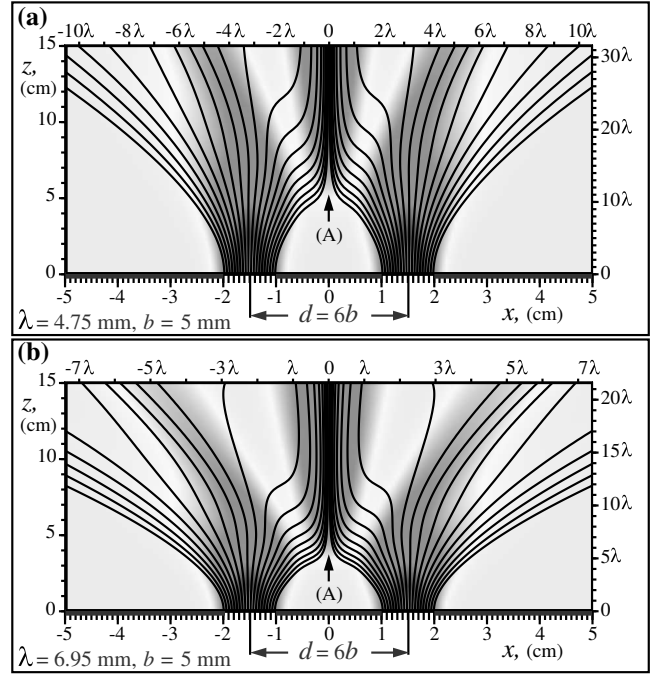}
  \end{picture}
  \caption{ 
  Bohmian trajectories going out from two slits~\citep{Sbitnev2009, Sbitnev2013b} are shown by black curves against the background of the probability density depicted by gray (light gray refers to low density; dark gray refers to high density):~(a) and~(b) relate to the bouncing droplets having wavelengths $\lambda = 4.75$~mm and 
$\lambda = 6.95$~mm~\citep{CouderForte:2006}, respectively. Arrows~(A) point to places where
 Bohmian trajectories dramatically change direction because of the existence of more than one slit.
  }
  \label{fig=6}
\end{figure}

Such a recoil of the fluid depends on preceding history of the droplet motion.
 It is the effect of the path memory stored in the Faraday waves~\citep{EddiEtAll:2011}. So, the droplet moves along an optimal path, along the Bohmian trajectory, with the velocity
\begin{equation}
  {\vec {\mathit v}}_0 = 
 {\rm Re}{{\Biggl\langle\Psi\Biggl|-{\bf i}\,{{\displaystyle\eta_{\sigma}}\over{\displaystyle m}}\,\nabla
 \Biggr|\Psi\Biggr\rangle}\over{\langle\Psi |\Psi\rangle}}
  = {{\nabla S}\over{m}}.
\label{eq=31}
\end{equation}
 Here $-{\bf i}(\eta_{\sigma}/m)\nabla$ is the velocity operator.

 It should be noted that the fluid medium should have a high susceptibility to any touching to its surface. It is achieved by generating the Faraday waves that are supported slightly below the supercritical threshold of the excitability. The closer the threshold the higher the susceptibility. Due to such a super-high susceptibility the Faraday waves are easily excitable long-lived waves. Reflecting from the environment boundaries they have a time for creating interference pattern around the droplet. The interference guides the droplet along an optimal path that is the Bohmian trajectory, Fig.~\ref{fig=6}.
 It is in good accordance with the de Broglie-–Bohm theory~\citep{DeBrogile1953, BohmHiley1982}
 where the wave function $\Psi$ plays a role of the pilot-wave guiding the particle along a most optimal trajectory.

\section{\label{subsec:level4}Physical vacuum as a superfluid}

 Let us define a material dependent parameter
\begin{equation}
   \hbar =e^{2}{{Z_0}\over{4\pi}}\alpha^{-1}\approx 1.0545\times10^{-34}~~{\rm J\cdot s}.
\label{eq=32}
\end{equation}
 Here $Z_0=(\mu_0/\epsilon_0)^{1/2}$ is the wave impedance of the free space ($\mu_0$ is the permeability of the free space, $\epsilon_0$ is its permittivity) and $e$ is the electron charge. A dimensionless parameter~$\alpha$ is the fundamental physical constant (the fine-structure constant) characterizing the strength of the electromagnetic interaction. 
 One can see that the parameter $\hbar$ is the reduced Planck constant.

 By substituting the constant $\hbar$ into Eq.~(\ref{eq=25}) instead of the parameter $\eta_{\sigma}$ we get the Schr\"{o}dinger equation describing flow of a peculiar fluid - the physical vacuum.
 The vacuum consists of pairs of particle-antiparticles. The pair per se is the Bose particle. These pairs stay at a temperature close to zero. 
It means, that the pairs make up Bose-condensate and, consequently, the vacuum represents a superfluid medium~\citep{Dirac1951, SinhaEtAl1976}. 
The superfluid medium represents a 'fluidic' nature of space itself. 
Another name for such an 'ideal fluid' is the aether~\citep{ Martins2012}. 

The physical vacuum (the aether), is a strongly correlated system dominated by collective effects~\citep{RobertsAndBerloff2001} with the viscosity equal to zero.
Nearest analogue of such a medium is the superfluid helium~\citep{Volovik2003}, which will serve us as an example for further consideration of this medium.
 The vacuum is defined as a state with the lowest possible energy. 
We shall consider a simple vacuum consisting of electron-positron pairs. The pairs fluctuate within the first Bohr orbit having energy about 
(13.6 eV)$\cdot 2\approx 27$ eV. Bohr radius of this orbit is $r_1\approx5.29\cdot10^{-11}$~m.
 These fluctuations occur about the center of their masses. The total mass of the pair, $m_p$, is equal to doubled mass of the electron, $m_e$. The charge of the pair is zero. The total spin of the pair is equal to 0. Whereas the angular momentum, $L$, is nonzero. For the first Bohr orbit  $L = \hbar$. The velocity of rotating about this orbit is $L/(r_1\cdot m_e)\approx 2.192•\cdot10^6$ m/s. So, it means that there is an elementary vortex. Ensemble of such vortices forms a vortex line.

 Now we may evaluate the dispersion relation between the frequency, $\epsilon(p) = \hbar\omega$, and wave number, $p=\hbar k$, as it done in~\citep{LeGan2013}. As follows from Eq.~(\ref{eq=25}) we have:
\begin{equation}
\label{eq=33}
 \epsilon(p) = {{1}\over{2m_p}}(p+p_Rf(p-p_R))^2.
\end{equation}
 Here $p_R=L/r_1=m_p {\mathit v}_R$ is the momentum of the rotation.
 The function $f(p-p_R)$ is a formfactor relating to the electron-positron pairs rotating about the center of their masses. The formfactor describes dispersion of the momentum $p$ around $p_R$ conditioned by fluctuations about the ground state with the lowest energy. The formfactor is similar to the Gaussian curve
\begin{equation}
   f(p-p_R) = \exp\Biggl\{
      -{{(p-p_R)^2}\over{2\sigma^2}}
                  \Biggr\}.
\label{eq=34}
\end{equation}
 Here $\sigma$ is the variance of this dispersion. And it is smaller or close to $p_R$.
 The dispersion relation~(\ref{eq=33}) is shown in Fig.~\ref{fig=7}. 
\begin{figure}
  \centering
  \begin{picture}(200,120)(10,30)
  \includegraphics{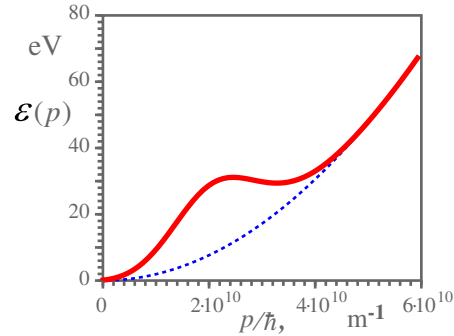}
  \end{picture}
  \caption{ 
The dispersion relation $\epsilon$ vs. $p$. The dotted curve shows the non-relativistic square dispersion relation $\epsilon\sim p^2$. The hump on the curve is a contribution of the roton component $p_Rf(p-p_R)$,  $p_R/\hbar\approx1.89\cdot10^{10}$ m$^{-}1$, and $\sigma = 0.5p_R$.  }
  \label{fig=7}
\end{figure}
The hump on the curve is due to the contribution of the rotating electron-positron pair about the center of masses. This rotating object is named the roton. The rotons are ubiquitous in the vacuum space. Motion of the roton in a free space is described by the following equation
\begin{equation}
 {\bf i}\hbar {{\partial\Psi}\over{\partial\,t}} =
 {{1}\over{2m_p}}(-{\bf i}\hbar + m_p{\vec {\mathit v}}_{_R})^2\Psi - C\Psi.
\label{eq=35}
\end{equation}
 Here $C\Psi$ determines an uncertain phase shift of the wave function, and most possible this phase relates to the chemical potential of a boson (the electron-positron pair)~\citep{LeGan2013}. We did not take into account contribution of this term in the dispersion diagram because of its smallness.

 As follows from the above computations Eq.~(\ref{eq=35}) can be reduced to the Euler equation:
\begin{equation}
  {{\partial\,{\vec {\mathit v}}_R}\over{\partial\,t}}
  + [{\vec\omega}\times{\vec {\mathit v}}_R] 
 = - {{\nabla P}\over{m_p\rho}},
\label{eq=36}
\end{equation}
 describing a flow of the inviscid incompressible fluid under the pressure field $P$. One can see from here, that the Coriolis force appears as a restoring force, forcing the displaced fluid particles to move on circles. The Coriolis force is the generating force of waves called ''inertial waves''~\citep{LeGan2013}. So the Euler equation admits a stationary solution for uniform swirling flow under the pressure gradient along $z$. 

The twisted vortex states observed in superfluid $^3$He-B~\citep{EltsovEtAl2006} are closely related to the inertial waves in rotating classical fluids. The superfluid initially is at rest. The vortices are nucleated at a bottom disk platform rotating about axis~$z$~\citep{LounasmaaAndThuneberg1999}, see Fig.~\ref{fig=8}. The platform is in the normal state. The Coriolis forces take part in twisting of the vortices. The twisted vortices grow upward along the cylinder axis~\citep{LounasmaaAndThuneberg1999}. 
\begin{figure}
  \centering
  \begin{picture}(200,220)(40,20)
  \includegraphics{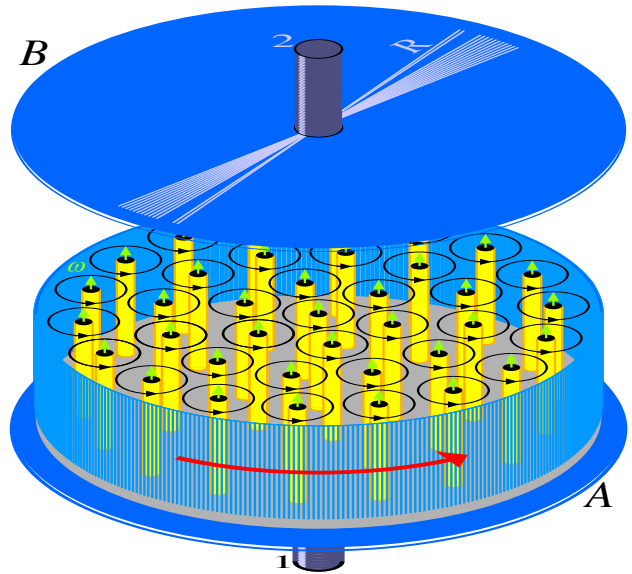}
  \end{picture}
  \caption{ 
Rotation of the superfluid is not uniform but takes place via a lattice of quantized vortices, whose cores (yellow) are parallel to the axis of rotation~\citep{LounasmaaAndThuneberg1999}. Green arrows are the vorticity $\omega$. Small black arrows indicate the circulation of the superfluid velocity $u_R$ around the cores. The vacuum is between two nonferromagnetic disks, {\it A} and {\it B},  fixed on center shafts, 
{\bf 1} and {\bf 2}, of electromotors~\citep{Samohvalov2013}. Radiuses of the disks are $R = 82.5$ mm and distance between them can vary from 1 to 3 mm and more. The vortex bundle rotates rigidly with the disk {\it A}. As soon as the vortex bundle reaches the top disk {\it B} it begins rotation as well.}
  \label{fig=8}
\end{figure}
 Samohvalov has shown through the experiment~\citep{Samohvalov2013}, that the vortex bundle induced by rotating the bottom nonferromagnetic disk {\it A} leads to rotation of the top nonferromagnetic motionless disk {\it B}, Fig.~\ref{fig=8}.
 Both disks at room temperature have been placed in the container with technical vacuum at 0.02 Torr.
 The utmost number of the vortices that may be placed on the square of the disk {\it A} is about $N_{\max} = (2\pi R^2)/(2\pi r_1^2) = 2\cdot10^{18}$, where $R = 18.5$ mm is the radius of the disk and $r_1 \approx 5.29\cdot10^{-11}$ m is the radius of the first Bohr orbit.
 Really, the number of vortices situated on the square, $N$, is considerably smaller. It can be evaluated by multiplying $N_{\max}$ by a factor $\delta$. This factor is equal to the ratio of the geometric mean of the velocities 
${\mathit v}_R = \hbar/(r_1\cdot m_e) \approx 2.192\cdot10^6$ m/s and $V_D = R\Omega$ to the arithmetic mean of these velocities. Here $\Omega$ is an angular rate of the disk {\it A}. So, we have
\begin{equation}
   N = N_{\max}{{\sqrt{{\mathit v}_R\cdot V_D}}\over{{\mathit v}_R + V_D}}
  = N_{\max}\sqrt{{{V_D}\over{{\mathit v}_R}}} \approx 6\cdot10^{15}.
\label{eq=37}
\end{equation}
 At the angular rate $\Omega = 160$~1/s~\citep{Samohvalov2013} the disk velocity $V_D = 13.2$ m/s. 
Now we can evaluate the kinetic energy of the vortex bundle induced by the rotating disk {\it A}. 
This kinetic energy is $E = N\cdot m_p {\mathit v}_R^2/2 \approx 0.026$~J. It is sufficient for transfer of the moment of force to the disk {\it B}. Measured in the experiment~\citep{Samohvalov2013} the torque is about 0.01 N$\cdot$m. 
So, the disk {\it B} can be captured by the twisted vortex bundle to be rotated.
 
 Light traveling along two paths through the space between the disks suffers a phase shift~\citep{Martins2012} 
 as it is in the famous experiment of~\citet{AharonovBohm1959}.

\section{\label{subsec:level5}Conclusion}

The Navier-Stokes equation contains terms describing existence of vortex motions in the fluid media.
Under some  conditions the Navier-Stokes equation can be reduced to the Schr\"{o}dinger-like equation. This equation can contain terms describing both the irrotational solutions and vortex ones. The latter solutions are due to appearance of a term in the kinetic momentum operator responsible for the vortex motion. This term looks as the vortex velocity multiplied by mass of the particle. The Helmholtz theorem gives a basis for such an inclusion of this term. 
The inclusion is similar to that of the magnetic vector potential to the kinetic momentum operator~\citep{Martins2012}.

Due to such an inclusion the Schr\"{o}dinger equation admits emergence of solutions like vortex balls moving along optimal paths called as the Bohmian trajectories. Works of the French scientific team~\citep{CouderForte:2006, CouderEtAll:2005, ProtierEtAl:2006, EddiEtAll:2011} throw light on dynamics of such a motion. The vortex ball rolls, by bouncing from one plot to another, as it moves along some path. Because of such bounces the ball induces weak long-lived waves on the fluid surface. The waves form an interference pattern guiding the ball along the path due to creating constructive and destructive interference ahead. This observation supports the de Broglie Bohm's theory about the guiding wave function.

Returning to the quantum realm we find that the Schr\"{o}dinger equation describes motion of a special fluid - the physical vacuum as a superfluid fluid presented by self-organized Bose ensemble which consists of virtual particle-antiparticle pairs. It confirms insights of Madelung~\citep{Madelung1926} and Bohm with Vigier~\citep{BohmVigier1952} about a fluid-like quantum medium. 
This medium, called also the aether, has mass and is populated by the particles of matter which exist in it and move through 
it~\citep{PinheiroBuker2012, Martins2012, SaravaniEtAl2013, Sola2013}.
The particle traveling through that medium perturbs virtual particle-antiparticle pairs which, in turn, create both constructive and destructive interference ahead the particle. Thereby the virtual pairs provide an optimal, Bohmian path for the particle. It is interesting to notice that the vacuum undergoes fluctuations in the ground state that have wave-like nature. These wave-like fluctuations are akin the subcritical Faraday oscillations owing to which motion of the droplet along optimal, Bohmian path can be attainable.

Particle-antiparticle pairs composing vacuum are on the lowest ground states and fluctuate near the first Bohr orbit. This orbit has the angular momentum $L = \hbar$. Sum of all angular momenta in the case of their alignment within a container containing two disks (one is rotating) is sufficient in order to initiate a torque in the disk initially unmoved. 
Initiation of the torque is observed only in the vacuum, but not at normal atmosphere~\citep{Samohvalov2013}. 
It occurs owing to self-organization of the vortex lines into the vortex bundles growing from the bottom rotating disk up to the upper unmoved disk~\citep{EltsovEtAl2006, LounasmaaAndThuneberg1999}.\\

\begin{acknowledgments}

 The author thanks prof.~Robert~Brady for valuable propositions.
 The author would like to thank Mike Cavedon for useful and valuable remarks and offers.

\end{acknowledgments}

%=

\bibliographystyle{jfm}

% -----------------------------------------------------------

\end{document}